\documentclass[conference]{IEEEtran}
\IEEEoverridecommandlockouts
\usepackage{algorithm}
\usepackage{algpseudocode}
\usepackage{array}
\usepackage{adjustbox}
\usepackage[caption=false,font=normalsize,labelfont=sf,textfont=sf]{subfig}
\usepackage{textcomp}
\usepackage{stfloats}
\usepackage{url}
\usepackage{verbatim}
\usepackage{graphicx}
\usepackage{cite}
\usepackage{longtable}
\usepackage{lipsum} 
\usepackage{amsmath,amssymb,amsfonts}
\usepackage{xcolor}
\usepackage{multirow}
\usepackage{subfig}
\usepackage{makecell}
\usepackage{enumitem}
\usepackage{derivative}
\usepackage{balance}

\usepackage[utf8]{inputenc}
\usepackage{float}
\usepackage{afterpage}
\usepackage{tikz}
\usepackage{diagbox}
\def\BibTeX{{\rm B\kern-.05em{\sc i\kern-.025em b}\kern-.08em
    T\kern-.1667em\lower.7ex\hbox{E}\kern-.125emX}}
\begin{document}

\title{SPERO: \underline{S}imultaneous \underline{P}ower/\underline{E}M Side-channel Dataset Using \underline{R}eal-time and \underline{O}scilloscope Setups
\thanks{This work has been supported in part by the US Army Research Office (ARO) under award number W911NF-19-1-0102.}
}

\author{Yunkai Bai\\
\IEEEauthorblockA{
University of Florida\\
Gainesville, FL, USA \\
baiyunkai@ufl.edu}
\and
\IEEEauthorblockN{Rabin Yu Acharya}
\IEEEauthorblockA{
Intel Corporation\\
Hillsboro, OR, USA \\
rabin.yu.acharya@intel.com}
\and
\IEEEauthorblockN{Domenic Forte}
\IEEEauthorblockA{
University of Florida\\
Gainesville, FL, USA \\
dforte@ece.ufl.edu}
\and

}

\maketitle

\begin{abstract}

Cryptosystem implementations often disclose information regarding a secret key due to correlations with side channels such as power consumption, timing variations, and electromagnetic emissions. 
Since power and EM channels can leak distinct information, the combination of EM and power channels could increase side-channel attack efficiency. 
In this paper, we develop a miniature dual-channel side-channel detection platform, named RASCv3 to successfully extract subkeys from both unmasked and masked AES modules. 
For the unmasked AES, we combine EM and power channels by using mutual information to extract the secret key in real-time mode and the experiment result shows that less measurements-to-disclosure (MTD) is used than the last version (RASCv2). Further, we adopt RASCv3 to collect EM/Power traces from the masked AES module and successfully extract the secret key from the masked AES module in fewer power/EM/dual channel traces. 
In the end, we generate an ASCAD format dataset named SPERO, which consists of EM and power traces collected simultaneously during unmasked/masked AES module doing encryption and upload to the community for future use.
\end{abstract}


\section{Introduction}
Modern cryptosystems operate through semiconductor logic gates, which are constructed out of transistors. 
As data is encrypted or decrypted, these gates emit electromagnetic (EM) radiation and cause variations in power due to the transition of these transistors. 
This characteristic has been leveraged over the past few decades to extract secret keys from cryptosystems in so-called side-channel attacks (SCAs) using power and EM.

The concept of power-based side-channel attacks was first introduced in Paul Kocher's seminal paper\cite{kocher1999differential}. 
Kocher et al.\cite{kocher1999differential,kocher2011introduction, kocher1998introduction} presented the differential power analysis (DPA) attack which exploits and enlarges power leakage differences when DES cryptographic systems process various logic bits. 
In 2004, Brier and colleagues\cite{brier2004correlation} developed the correlation power analysis (CPA) technique, which utilizes the linear relationship between power consumption and its hamming distance (HD) to identify the correct secret key from all possible hypotheses. 
The EM-based side-channel attack originated from\cite{gandolfi2001electromagnetic} in 2001 where Gandolfi et al. successfully recovered keys from EM leakage on three different CMOS chips. 
Agrawal et.al\cite{agrawal2003side} demonstrated EM vulnerabilities in basic implementations of widely used cryptographic systems like DES\cite{pub1999data}, RSA\cite{zhou2011research,mahajan2013study} and COMP128\cite{zhou2013need} across devices such as smart cards, cryptographic tokens, and SSL accelerators. 
Furthermore, Ding et al.\cite{ding2009correlation} presented the correlation electromagnetic attack (CEMA) on the P89C668 microcomputer. 

Besides these single-channel SCA attack methods, multiple papers\cite{souissi2012towards,standaert2008using,bai2023dual} turn to recover secret keys from dual-channel information, e.g. using both EM and power traces. 
Compared with single-channel SCAs, dual-channel SCAs contain more information, and therefore, increase the SCA success rate with fewer traces.
Standaert et al.\cite{standaert2008using} concatenated EM and power traces and their experiments show less entropy is achieved after concatenation. 
Souissi et.al\cite{souissi2012towards} try to combine multiple channels by summing up the square of each channel. 
Compared with the single power channel, this sum of squares method has a larger signal-to-noise ratio (SNR), and the experiments demonstrate that fewer traces are needed to attack the same SBox. 
However, dual-channel attacks also require more processing resources and lack granularity in the combination of features at different time indexes. 
Furthermore, nearly all SCA attack papers, either single-channel or dual-channel attack, are limited to offline setups where data collection relies on traditional instruments such as oscilloscope and commercial EM probe, and processing relies on PC platforms. 
The offline mode faces various challenges in practical scenarios outside of the lab environment, such as power constraints, tiny space, on-board processing and remote communication. 
To solve these problems, Bai et.al\cite{bai2023dual} combine EM and power channels in a linear fashion by using mutual information\cite{peng2005feature,ding2005minimum,cover1999elements} to determine the optimal coefficients for each feature. 
The proposed methodology is also implemented onto a miniature side-channel platform named RDCP (or RASCv2) to simultaneously measure dual-channel traces and process them to extract AES subkeys in real time. 
The experiment result shows the success rate of dual-channel increases by at least 30\% compared to single power/EM channels in the offline mode and over 50\% in the real-time mode. 

In this paper, we upgrade RASCv2 to its third version, RASCv3, and create the first open-source SCA dataset containing both power and EM measurements.
Our main contributions are summarized as follows:  

\begin{itemize}[leftmargin=*]

\item RASCv3’s design is described and its features are compared with other side channel measurement systems, including RASCv2, traditional oscilloscope and probe, and Chipwhisperer. 
Unlike the oscilloscope and the Chipwhisperer, RASCv3 has advantages in low cost and diminutive size while obtaining desired results.

\item We also upgrade and describe RASC's near-field antenna design so that EM traces collected with RASCv3 have higher SNR.

\item RASCv3’s offensive capabilities are compared with RASCv2 for \textit{unmasked} AES encryption in single and dual-channel variants. 
Fewer traces are needed to achieve a 100\% subkey extracting rate. 
Furthermore, \textit{masked} AES is proven breakable by RASCv3 for all 16 subkeys with more traces needed than unmasked cases. 
For this, we extend a second-order side-channel attack to utilize information from both power and EM channels.

\item The dual-channel traces collected by using oscilloscope and RASC setups are named SPERO, and have been uploaded to GitHub for academic use by peers\cite{SPERO}. 
To our knowledge, there is no other dataset collecting EM and power simultaneously.

\end{itemize}

The remainder of this paper is organized as follows. 
Section~\ref{sec:overview} introduces the design of RASCv3, its upgrade from RASCv2, and its comparison with other side-channel detection setups. 
Section~\ref{sec:antenna} introduces the key concepts of designing a near-field antenna for RASCv3 and its comparison with commercial EM probe\cite{EMV} and RASCv2's antenna. 
In Section~\ref{sec:experiment}, we implement the proposed methodology into RASCv3 and extract subkey from both normal (unmasked) and masked AES encryption.
It includes the experimental setup, results, and discussion. We conclude and offer directions for future work in the last section.

\section{Overview of RASC}\label{sec:overview}

\subsection{What is RASC?}
Traditional side-channel analysis turns to sophisticated instruments such as oscilloscopes which are infeasible for in-field use and power hungry. 
To resolve this, we previously proposed RASC (short for, remote access to a side-channel platform), an external miniature monitor, that performs side-channel analysis for offensive (e.g., secret extraction) and defensive purposes (e.g., disassembly and malware detection) on a target chip. 
The first version of RASC was introduced by Stern et al. in 2019\cite{stern2019rasc}, and the upgrade version (RASCv2) was first presented by Bai et al. in 2022\cite{bai2022rascv2}. 
In this section, we introduce the design of the third version of RASC (RASCv3) and compare it with RASCv2 and traditional side-channel instruments, such as oscilloscope and EM probe. 

\subsection{Structure and Specs of RASCv3}

\begin{figure}
  \centering
    \includegraphics[width=0.4\textwidth]{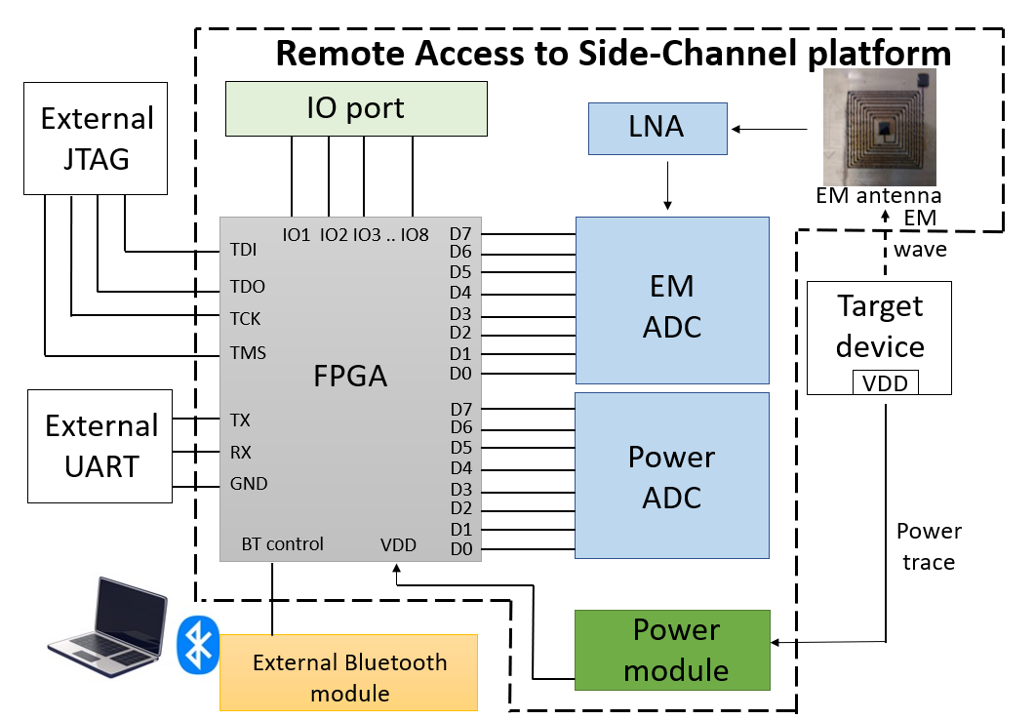}
     \caption {RASCv3 schematic.}
     \label{RASCschematic}
\end{figure}

Figure~\ref{fig:RASCV3structure} presents the structure of the RASCv3 while its high-level schematic is shown in Figure~\ref{RASCschematic}.
RASCv3 consists of two boards, referred to as PCB1 and PCB2. 
PCB1 contains two ADCs for simultaneously sampling EM and power traces, an FPGA for data processing, and a few I/O ports for external connections. 
PCB2 has an electrical pad connecting to an external printable near-field antenna, and an amplifier for enhancing captured EM waves. 
The connection of two PCB boards is presented in Figure~\ref{fig:RASCV3structure}, i.e., the two boards could be attached to each other and stacked on top of the target chip.
For repeatability, we 3D print a holder to position  RASC\cite{bai2022rascv2} on top of the target chip in its operational scenario. 
EM traces are could be gathered by the near-field of antenna and the power of the RASC board could be supplied by the power source of the target chip. 
Compared with RASCv2, RASCv3 adjusts its design and upgrades its key specs so that it can handle more sophisticated experiments and practical applications. 
The main chip and functional upgrades are:

\begin{figure}
  \centering
    \includegraphics[width=0.4\textwidth]{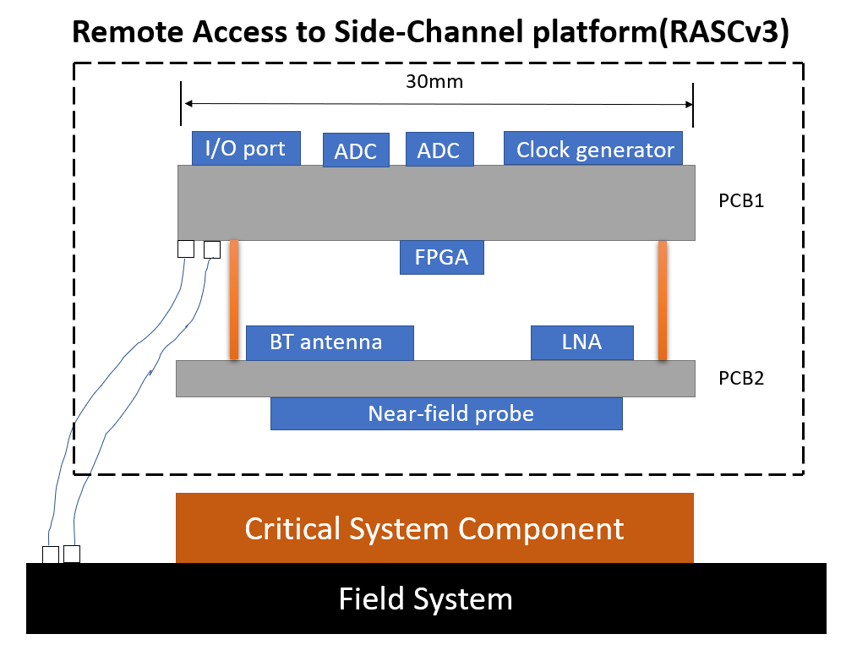}
     \caption {RASCv3 cross-sectional structure and CONOP.}
     \label{fig:RASCV3structure}
\end{figure}

 \begin{enumerate}
      
 \item \textbf{ADCs.} Compared with the 8-bit ADC(ADC08200) on RASCv2 (200MS/s, 500MHz, 8-bit), the new ADC on RASCv3 (LT2242\cite{LT2242}) could achieve \textit{faster sampling} (250 MS/s at maximum), \textit{larger bandwidth} (1GHz), and \textit{higher voltage sensitivity} (12-bit). Faster sampling allows the ADC to capture more data points within a given period, and it recovers side-channel leakage more accurately as evident in later experiments. With 12-bit resolution, the minimum detecting voltage is around 1mV, allowing RASCv3 to infer minor changes in power/EM. Also, RASCv2 output only support CMOS mode so that its detecting range is from 0V to 3V. Thus, it lacks the ability to detect negative voltage and this seriously affects the integrity of EM traces\cite{bai2022rascv2}. In RASCv3, the LT2242 supports LVCMOS mode so that its voltage detecting range is from -1V to 1V, allowing entire EM traces to be collected.
 
\item \textbf{FPGA.} Compared with the Spartan 3e FPGA\cite{Spartan3e} on RASCv2, Artix-7 xca100t\cite{Artix7} is chosen on RASCv3 for its \textit{larger memory}, \textit{higher clock rate}, and additional \textit{\textbf{}I/O ports}. RASCv2's memory was sufficient for simple algorithms such as linear SVM classifiers with limited features. In RASCv3, the memory is over 1000Kb, allowing more complex and nonlinear classifiers such as QDA. These upgrades allow RASCv3 to contain more sophisticated SCA methods and the ability to process longer traces in real-time experiments.
  
 \item \textbf{EM Antenna.} In RASCv2, we inserted a four-loop internal near-field antenna into PCB2. It helps gather EM waves from the target board, and the subkey could be successfully extracted in an unmasked AES-128 subkey extracting experiment \cite{bai2022rascv2}. However, as mentioned in \cite{bai2022rascv2}, the SNR ratio of the internal layer of the antenna is affected by the connection between layers, and this impacts key extraction efficiency. In RASCv3, we substitute the internal antenna with a cheaper and printable near-field antenna. This new  antenna has advantages in \textit{reduced cost} ($<$ \$1), \textit{higher SNR ratio}, and \textit{better portability}.

 \end{enumerate}

\begin{figure}
  \centering
    \includegraphics[width=0.4\textwidth]{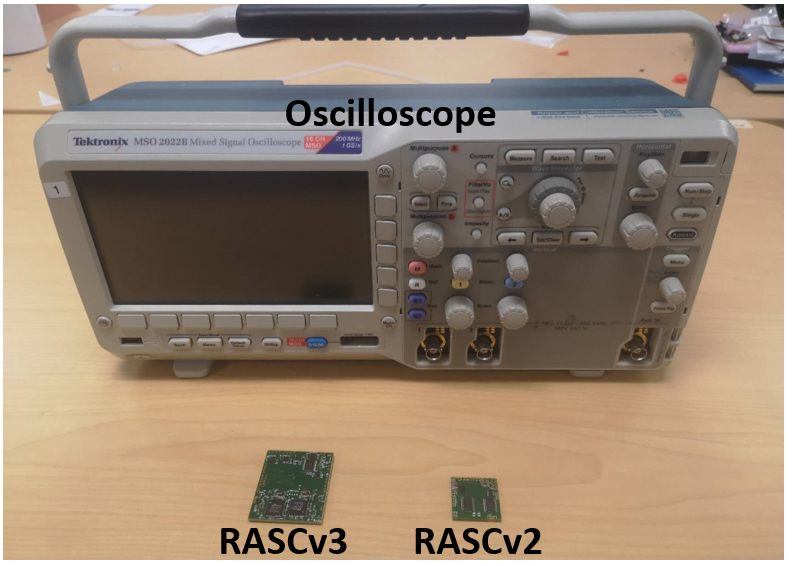}
     \caption {Size comparison of oscilloscope, RASCv2, and RASCv3.}
     \label{RASCcomparison}
\end{figure}

 Figure~\ref{RASCcomparison} shows RASCv3 compared to RASCv@ and an oscilloscope. 
 The size of RASCv2 is about this size of a quarter while RASCv3 is about 50\% larger.
 Considering the balance of size and performance, RASCv3 could still be thought of as small and suitable for deployment outside of lab environments.

\subsection{Working Modes of RASC}

In this paper, we consider offline and real-time modes to extract AES key information:
\begin{enumerate}

\item \textbf{Offline mode.} Here, an oscilloscope and commercial probe are used to collect EM/power traces and store them first. These traces are sent 
 to a PC through UART for processing and key extraction. 

\item \textbf{Real-time mode.} In this mode, we use RASC to collect power/EM traces and implement DPA attacks on RASC's FPGA to extract the subkey bits internally. The extracted subkey bits are then transmitted to a PC through UART. 
\end{enumerate}


\subsection{Comparison between RASC and oscilloscope} \label{sec:RASCvScope}

Table~\ref{Comparision} compares RASCv3, RASCv2, MDO3102 oscilloscope, and ChipWhisperer~\cite{Chipwhisperer} which is a popular commercial board. The highlights are as follows:

\begin{enumerate}

\item\textbf{Price.} The traditional side-channel analysis system consists of a commercial EM probe and oscilloscope. The oscilloscope (Tektronix MDO 3102) in our lab can sample data over 5 GS/s and costs over \$16,000, and the commercial EM probe also costs over \$800. However, the total cost of RASCv2 is around \$250 (produced at low volume), which is comparable to ChipWhisper lite edition (produced at high volume). RASCv3 upgrades RASCv2's functionality and costs around \$400 (also at low volume). The cost of RASC (v2 and  v3) could be dropped to a much lower value if RASC is fabricated in larger quantities like commercial systems.

\item\textbf{Size.} RASCv2 and RASCv3's size are shown in Figure~\ref{RASCcomparison}. Even though RASCv3 is 50\% larger in size, it still could be thought of as small. The tiny body of two versions of RASC could let them be easily placed into narrower spaces to monitor IoT devices. Benchtop oscilloscopes commonly used in SCA cannot.

\item\textbf{Voltage Testing Range.} The testing voltage range of two ADCs on RASCv2 is from 0V to 3.3V, and RASCv3 is from -1V to 1V. For both power/EM ADC, the range of RASCv3 is acceptable since this range covers the power supply of most commercial FPGA/MCU development boards. Compared with two versions of RASC and ChipWhisperer, the oscilloscope has a much more extensive testing range. 

\item\textbf{Sampling Speed.} The oscilloscope has the advantage of sampling speed. As for Tektronix MDO 3102, the maximum sampling speed could reach 5 GS/s. The sampling speeds of RASCv2/RASCv3 and ChipWhisperer are smaller than the oscilloscope. Sampling speed is more important in sophisticated experiments such as fine-grained malware detection and disassembly. 
 
\item\textbf{Remote Communication.} The RASCv2 and RASCv3 both support Bluetooth communication over 20 meters. This functionality allows RASC to work remotely and communicate results to attack/defense parties. However, ChipWhisper and the oscilloscope do not support this feature.

\item\textbf{Resolution.} RASC's resolution is increased from 10mV to 1mV between versions 2 and 3. The new version of RASC (RASCv3) is comparable to the resolution of the ChipWhisperer board. Even though the oscilloscope can detect faint changes in voltage/EM, RASC performs quite well in our experiments.

\item\textbf{Programmable.} On both RASCv2 and RASCv3 PCB1, we set 14 I/O ports. FPGA code can be loaded on PCB1 to achieve many different functionalities. The oscilloscope can access Matlab on the laptop and supports API code. However, this is not as flexible as compared to an FPGA. 

\end{enumerate}

In conclusion, RASCv3 has advantages in low price, small footprint, and remote communication. At the same time, compared with oscilloscopes, it has limitations in the sampling speed, resolution, and voltage testing range. However, in many scenarios like cracking AES-128 subkeys, there is no need to sample traces at such high speeds. In other words, the RASC can serve as a substitute for an oscilloscope in practical defense/offense applications.

\begin{table*}[t]\setlength{\tabcolsep}{4pt}
\scriptsize
\centering
  \caption{Comparison between traditional side-channel analysis systems, ChipWhisperer, RASCv2, and RASCv3.}

  \begin{tabular}{|c|c|c|c|c|}
\hline
     \textbf{Property\textbackslash Setup} & \textbf{Oscilloscope\cite{MDO3102} + EM antenna\cite{EMV}} & \textbf{ChipWhisperer-Lite 32-Bit\cite{Chipwhisperer}} & \textbf{RASCv2\cite{bai2022rascv2}} & \textbf{RASCv3 (This paper)}\\
\hline
    \textbf{Cost} &  $>$\$16,000 & \$250 & \$250&  \$400  \\
\hline    
    \textbf{Size} & 20cm$\times$42cm$\times$15cm & 11.5cm$\times$8.8cm & 2.5cm$\times$2.5cm & 3.8cm$\times$3.8cm\\
\hline
    \textbf{Test Voltage Range} & [-20V, 20V] & [-1V, 1V] & [0, 3.3V] & [-1V, 1V]\\
\hline
    \textbf{Sampling Speed} & 5 GS/s & 105 MS/s & 200 MS/s & 250 MS/s \\
\hline
    \textbf{Remote Communication} & No & No & Yes & Yes \\
\hline
    \textbf{Resolution} & 16-bit, 60$\mu$V & 10-bit, 1.95mV & 8-bit, 10mV & 12-bit, 0.5mV \\   
\hline
    \textbf{Programmable} & No & Yes & Yes & Yes\\
\hline
  \end{tabular}
  \label{Comparision}
\end{table*}

\section{Near-field Antenna Design}\label{sec:antenna}

\subsection{Antenna Principles and Considerations}

In SCA real-time scenarios, the near-field antenna transforms EM leakage from the useful signal source (e.g., VDD or GND) on the target chip to current. After that, RASC detects the voltage when the current of the antenna goes through a load resistance. More power received by the antenna will cause a larger amplitude and higher SNR of EM traces. A higher SNR could increase the subkey extracting efficiency in the SCA experiment. Our near-field antenna is designed carefully to increase received power received power $P_r$ based on equations from a common radars book\cite{skolnik1980introduction}. 
\begin{equation}
P_{r}= \frac{P_{t}GA_{e}\sigma}
{(4\pi)^2R^4} 
\label{formula:Pt}
\end{equation}
Here, $P_{r}$ and $P_{t}$ stands for the received and transmitted power. $G$ is the antenna gain. $A$ is the effective aperture of the receiving antenna. $\sigma$ is the radar cross-section of the target. $R$ is the distance between the transmitter and the target. 

In our case, the diameter for the near-field antenna is decided by the size of the RASC's PCB2, and its max size is 2cm $\times$ 2cm. When working, the near-field antenna is arranged to a position with a fixed parallel angle and a fixed close distance (e.g., 1mm) to a fixed signal source on the target board. The transmitted power $P_{t}$, the distance $R$ and the directivity ($D$) could be thought of as the same. Besides, considering the diameter of our antenna, our antenna clearly receives the near-field EM wave, and we can ignore the radar cross-section ($\sigma$) since it is a far-field parameter. Thus, the parameters most significantly affecting the received power is the effective aperture ($A$) and the antenna gain ($G$). $A$ and $G$ are computed as follows:
\begin{equation}
A_{e}= e_{cd}\frac{\lambda^2}{4\pi}D
\label{formula:Ae}
\end{equation}
\begin{equation}
G= e_{cd}D
\label{formula:gain}
\end{equation}
In formula~\ref{formula:ecd}, $R_{r}$ is the radiation resistance, and $R_{L}$ is the loss resistance. The radiation resistance $R_{r}$ is used to represent in the receiving mode the transfer of energy from the free-space wave to the
where $D$ stands for the directivity of the antenna to the signal source and $e_{cd}$ is the radiation efficiency. The directivity of an antenna is defined as “the ratio of the radiation intensity in a given direction from the antenna to the radiation intensity averaged over all directions''\cite{balanis2016antenna}. Considering the angle between the near-field antenna and the signal source is set to a fixed ideal angle, the directivity remains the same. Thus, $A$ and $G$ are determined by the radiation efficiency ($e_{cd}$) as shown in Eqn.~\eqref{formula:ecd}.
\begin{equation}
e_{cd}= \frac{R_{r}}{R_{L}+R_{r}}
\label{formula:ecd}
\end{equation}
In the above equation, $R_{r}$ is the radiation resistance and $R_{L}$ is the loss resistance. The radiation resistance $R_{r}$ is used to represent the transfer of energy from the free-space wave to the antenna in the receiving mode\cite{balanis2016antenna}. The loss resistance $R_{L}$ refers to the resistance that results in power loss. $R_{r}$ and $R_{L}$ for N-turn circle antennas are presented as follows.
\begin{equation}
R_{r}= 20\pi^2(\frac{C}{\lambda})^4 N^2
\label{formula:RrN}
\end{equation}
\begin{equation}
R_{L}= \frac{NC}{2\pi b}\sqrt\frac{\omega\mu_{0}}{2\sigma}
\label{formula:RLN}
\end{equation}
Here, $C$ is the circumference, $N$ stands for the number of turns, $b$ is the diameter of the wires, $\omega$ is the angular frequency of the target source, $\mu_{0}$ is the permeability of free space, and $\sigma$ is the conductivity of the material from which the antenna is made. Substituting Eqns.~\eqref{formula:RrN} and \eqref{formula:RLN} into Eqn.~\eqref{formula:ecd}, we obtain
\begin{equation}
e_{cd}= \frac{20\pi^2(\frac{C^3}{\lambda^4})N}{\frac{1}{2\pi b}\sqrt\frac{\omega\mu_{0}}{2\sigma}+20\pi^2(\frac{C^3}{\lambda^4})N}.
\label{formula:ecdN}
\end{equation}
Based on Eqn.~\eqref{formula:ecdN}, it is clear the circumference ($C$) and turns ($N$) have a positive relationship with radiation efficiency ($e_{cd}$). 
In other words, under the case of the same distance and same direction, a larger size and/or more turns could lead to higher radiation efficiency ($e_{cd}$). 
This in turn could create a higher effective aperture ($A$) and the antenna gain ($G$) in order to receive more power from the signal source. 
However, increasing $C$ and $N$ does not mean everything for higher amplitude in the SCA experiments. 
The voltage received at the side-channel instrument (such as an oscilloscope or RASC) is presented in Eqn.~\eqref{formula:V}.
\begin{equation}
V_{inst}=\sqrt{\frac{P_{r}}{R_{ant}}}R_{inst}
\label{formula:V}
\end{equation}
$V_{inst}$ stands for the voltage at the side-channel instrument side. 
$R_{inst}$ denotes the load resistance at the side-channel instrument side and can be thought of as the same in the analysis. 
$P_{r}$ and $R_{ant}$ refer to the received power and resistance of the near-field antenna. 

Even if the radiation efficiency ($e_{cd}$) increases and the antenna gets more power ($P_{r}$), the more turns and larger size antenna could also significantly increase the resistance of the antenna $R_{ant}$. 
The induced current of the antenna could be seriously decreased, and the amplitude of the voltage of the instrument side could also be affected. 
For larger $V_{inst}$, the near-field antenna should receive much power ($P_{r}$) from the signal source but maintain a low resistance ($R_{ant}$) by adjusting the turns ($N$) and the circumference ($C$). 
Thus, the number of turns and size of the antenna should be adjusted carefully according to practical considerations, ensuring they are neither too large nor too small.

\begin{figure}
  \begin{minipage}{0.2\textwidth} 
    \centering
    \includegraphics[height=8cm, width=2.5cm]{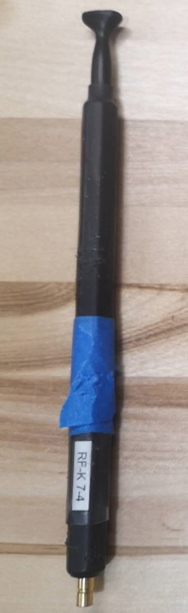} 
  \end{minipage}
\hspace{0.1cm}
 \begin{minipage}[c]{0.2\textwidth} 
 
  \begin{minipage}{0.2\textwidth} 
    \centering
    \includegraphics[height=2.2cm, width=2.2cm]{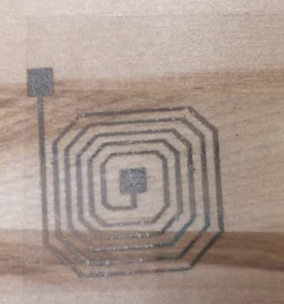}  \vspace{0.3cm}
  \end{minipage}  
  \\

  \begin{minipage}{0.2\textwidth} 
    \centering
    \includegraphics[height=2.2cm, width=2.2cm]{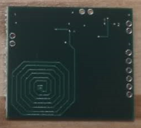}  \vspace{0.3cm}
  \end{minipage}  
  \\

  \begin{minipage}{0.2\textwidth} 
    \centering
    \includegraphics[height=2.2cm, width=2.2cm]{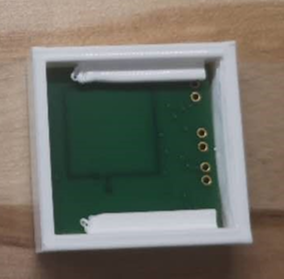} 
  \end{minipage}

  \end{minipage}
  \caption{Near-field antenna comparison. Left is the commercial EM probe, Right top is the printable antenna for RASCv3 board 3. The right middle is the near-field antenna on RASCv3 board 2. The right bottom is the internal antenna inside RASCv2 board 2. }
  \label{antennacomparison}
\end{figure}

\subsection{Antenna Designs Per RASC Version}

In RASCv1\cite{stern2019rasc}, a one-layer near-field antenna of ly 1.5cm$\times$1.5cm dimensions was inserted into the internal layer PCB2. Although it was supposed to obtain at least -40dB signal, it was unable to do so because PCB material in RASCv1 significantly weakened the received power ($P_{r}$) and caused low voltage amplitude ($V_{inst}$) at the instrument side. Furthermore, the low amplitude $V_{inst}$ led to poor SNR of the received EM waves. Thus, RASCv1's EM traces proved useless for SCA experiments.

Based on the lessons learned from RASCv1, the size of the near-field antenna in RASCv2 was increased to 2cm$\times$2cm, and 4 turns of the near-field antenna were separated into 4 internal layers of PCB2. The improvement of the circumference ($C$) and turns ($N$) significantly increased the detected voltage amplitude and the received EM traces were effective in the SCA experiments\cite{bai2022rascv2}. However, the transmission efficiency ($e_{cd}$) was still affected by the internal connection between PCB layers, and the structure of the internal layers increased RASCv2's cost. 


In RASCv3, we leave two connection pads on the bottom layer to connect to a 2cm$\times$2cm printable antenna shown in Figure~\ref{antennacomparison}. The printer used is the DMP-2850 from Dimatix. The ink used is NPS-L ink from Iwatani Corporation, and mainly consists of silver and isohexadecane. By printing the antenna, it is easy for us to iteratively improve its design and makes RASCv3 capable of adapting to more challenging scenarios. To prevent the short circuit between the near-field antenna and the signal source, we designed a holder (white frame presented in Figure~\ref{antennacomparison}). 

\begin{figure}
  \centering    \includegraphics[width=0.45\textwidth]{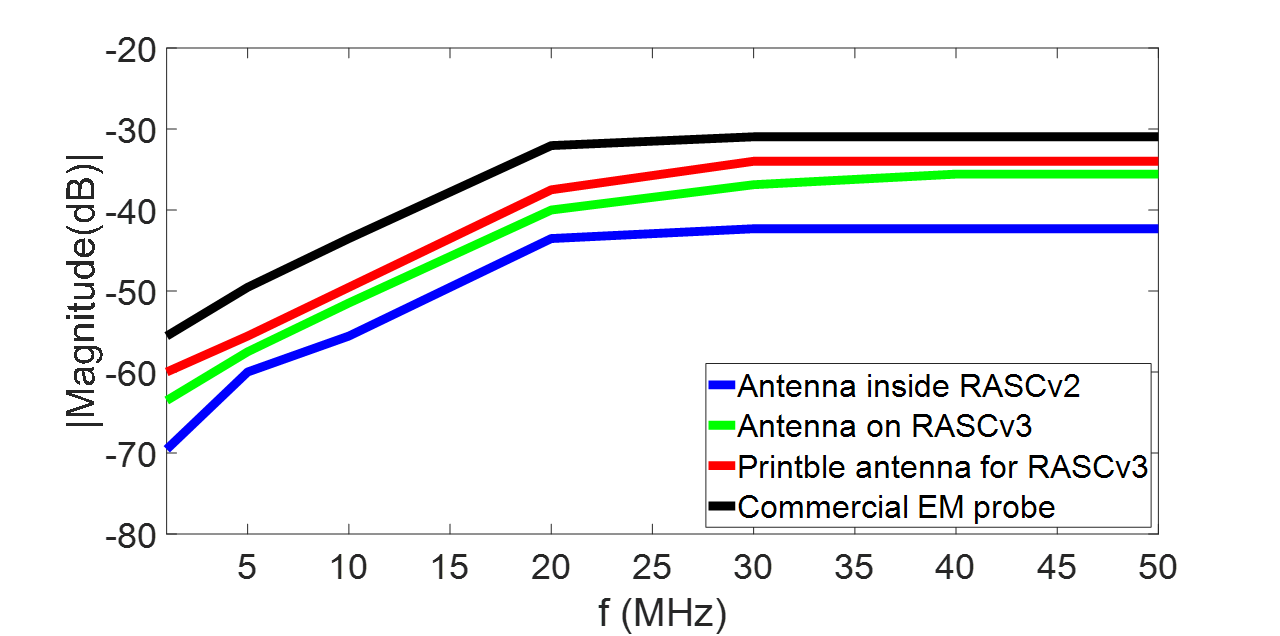}
     \caption {Near-field antenna magnitude comparison. }
     \label{magnitude_comparison}
\end{figure}

\subsection{Quantitative Antenna Comparison}\label{antenna:comparison}
Four different antennae are tested in this section: commercial EM probe (RF-K74 from LANGER EMV Technik\cite{EMV}), antenna inside RASCv2, antenna on RASCv3, and printable antenna for RASCv3. In figure~\ref{magnitude_comparison}, we test the magnitude of each versus frequency. All antennae are set to the same position 2mm above the Arduino UNO GND port during the response magnitude testing. Eqn.~\eqref{formula:db} explains how we calculate the magnitude. 
\begin{equation}
I=20\log \left (\frac{V_{test}}{V_{ref}} \right ) \label{formula:db}
\end{equation}
$V_{ref}$ stands for the reference voltage of the signal source and it is 3V in our experiment. $V_{test}$ denotes the response voltage of three antennae during the testing.

The EM response of the 4-turn internal antenna inside RASCv2 has the lowest magnitude among these four antennae. Luckily, we arrange 100$\times$ amplifier circuit on RASCv2 to augment the signal. The antenna on RASCv3 also has 4 turns but is designed on the bottom layer of RASCv3. Compared with the internal antenna inside RASCv2, it could receive signals without interference from PCB layers. Thus, the detected voltage from RASCv3 is higher than the internal antenna inside RASCv2. 
Compared with the antenna on the bottom of RASCv3, the printable antenna has 1.5$\times$ the circumstance and 2$\times$ wider traces. This modification lowers the resistance and could increase the magnitude of the printable antenna, as shown in Figure~\ref{antenna:comparison}. For the commercial EM probe, it has the highest magnitude among all four antennae in the testing. 

For the cost, the commercial EM probe is around \$800 while the internal antenna for RASCv2 is \$100. The printable antenna is the cheapest and it costs around 1 US dollars. Besides, our designed antennae (internal and printable) are more flexible to use. They can be situated on the target chip to work with either RASC or an oscilloscope. However, the commercial EM probe can only be matched with an oscilloscope and holder.


\section{Masked/Unmasked AES Cracking Experiments}\label{sec:experiment}

\subsection{Introduction to Masked and Unmasked AES} \label{sec:2orderattack}

The Advanced Encryption Standard (or AES)\cite{AES} is a widely used symmetric-key encryption algorithm that provides strong security for data protection. While it remains effective against brute force, mathematical, and even quantum attacks, it is still vulnerable to side-channel attacks (SCAs). SCAs utilize the side-channel leakage caused by transitions inside gates of a chip to infer secret information of the encryption module, such as encryption keys. Till now, many papers have shown that SCAs could extract subkeys of AES with few traces. 

\begin{algorithm}\label{algorithm:masked}

\small

\caption{Masked AES-128 implementation}
\begin{algorithmic}[1]
\State Input: 16-byte plaintext $p[0], \ldots, p[15]$
\State \ \ \ \ \ \ 16-byte mask vector $m[0], \ldots, m[15]$
\State \ \ \ \ \ \ 16-byte master key $s[0], \ldots, s[15]$
\State Output: 16-byte ciphertext $c[0], \ldots, c[15]$

\For{$i = 0$ to $3$} \Comment{Mask RoundKey}
    \State $s'[i*4] = s[i*4] \oplus (m[6] \oplus m[4]);$
    \State $s'[i*4+1] = s[i*4+1] \oplus (m[7] \oplus m[4]);$
    \State$s'[i*4+2] = s[i*4+2] \oplus (m[8] \oplus m[4]);$
    \State $s'[i*4+3] = s[i*4+3] \oplus (m[9] \oplus m[4]);$
\EndFor

\For{$i = 0$ to $3$} \Comment{Mask Plaintext}
    \State $p'[i*4] = p[i*4] \oplus (m[6] \oplus 0);$
    \State $p'[i*4+1] = p[i*4+1] \oplus (m[7] \oplus 0);$
    \State $p'[i*4+2] = p[i*4+2] \oplus (m[8] \oplus 0);$
    \State $p'[i*4+3] = p[i*4+3] \oplus (m[9] \oplus 0);$
\EndFor

\For{$i = 0$ to $15$} \Comment{Mask Sbox}
    \State $sbox'[i \oplus m[4]] = sbox[i] \oplus m[5];$
\EndFor
\For{$round = 1$ to $9$}
\For{$i = 0$ to $15$}
    \State $state[i] = p'[i] \oplus s'[i]$;
\EndFor
\For{$i=0$ to $15$}
    \State $state[i] = p'[i] \oplus s';$
\EndFor
\For{$i=1$ to $16$}
    \State $state[i] = sbox[state[i]];$
\EndFor
\State ShiftRows($state$);
\State Remask($state$, $m[0]$,$m[1]$,$m[2]$,$m[3]$,$m[5]$);
\State MixColumns($state$);
\EndFor
\State $\vdots$ \hfill\Comment{Last round}
\State \Return $(c[1], \ldots, c[16])$
\end{algorithmic}
\end{algorithm}

Masked AES, however, is a variant of AES designed to enhance the security of AES implementations, particularly against SCAs. Attacking a first-order masking scheme requires a second-order SCA. The masked AES code used in this paper is available on Github\cite{maskedAES}, and is listed in Algorithm~\ref{algorithm:masked} for the reader's reference. Unlike normal (unmasked) AES encryption, masked AES randomly generates 16 8-bit masked vectors ($m[0],..m[15]$) to mask all processed values, including round key, plaintext, and the Sbox. In line 7 to 10 of Algorithm~\ref{algorithm:masked}, $m[4]$, $m[6]$, $m[7]$, $m[8]$, and $m[9]$ are used to mask 16 round keys. In line 14 to 17, $m[6]$, $m[7]$, $m[8]$, and $m[9]$ are adopted to mask 16 plaintext bytes. In lines 20-22, the Sbox is masked with $m[5]$. The intermediate value after AddRoundKey and Sbox operation are masked by $m[4]$ and $m[5]$ in Eqns.~\eqref{formula:HWmaskedAES1} and~\eqref{formula:HWmaskedAES2}.\footnote{Apostrophes in equations and in Algorithm~\ref{algorithm:masked} refer to masked variables.}  After masking, first-order attacks cannot make the correct assumption of any intermediate values after masking AddRoundKey and Sbox outputs.
\begin{equation}
\small
p'[i] \oplus s'[i] = p[i] \oplus s[i] \oplus m[4], i=1,\hdots,16
\label{formula:HWmaskedAES1}
\end{equation}
\begin{equation}
\small
sbox'(p'[i] \oplus s'[i]) = sbox(p[i] \oplus s[i]) \oplus m[5], i=1,\hdots,16
\label{formula:HWmaskedAES2}
\end{equation}
Though one cannot directly guess the output of a single Sbox output, we still can guess the XOR of two Sbox outputs since all 16 Sbox bits share the same mask ($m5$), and XOR of any two Sbox output bits could remove the random mask ($m5$)\cite{messerges2000using,oswald2006practical}. This is presented in Eqn.~\eqref{formula:HWdifference_masked1}. Here, $sbox$ is unmasked Sbox and $sbox'$ is masked Sbox. $state$ is the intermediate value during the encryption.
\begin{align}\label{formula:HWdifference_masked1}
&HW(sbox'(state[i]) \oplus sbox'(state[j])) \nonumber
\\&=HW(sbox(state[i]) \oplus sbox(state[j]))
\end{align}
If we only focus on a single bit of XOR value of two masked Sbox outputs, the hamming weight (HW) of XOR value also equals to the absolute difference of two masked Sbox output HW\cite{oswald2006practical}, i.e.,
\begin{align}\label{formula:HWdifference_masked2}
&HW(sbox'(state[i]) \oplus sbox'(state[j])) \nonumber
\\&=HW(sbox(state[i]) \oplus sbox(state[j])) \nonumber
\\&=||HW(sbox'(state[i])|-|HW(sbox'(state[j]))||
\end{align}

When a second-order attack is implemented to attack this masking scheme, two 8-bit subkeys are attacked at the same time, and thus 65536 (256$\times$256) hypotheses need to be checked. Based on Eqn.~\eqref{formula:HWdifference_masked2}, we pair two subkeys from all 16 subkey bits ($s[0]$ and $s[1]$,$s[2]$ and $s[3]$, $\hdots$, $s[14]$ and $s[15]$) and guess two subkeys of masked AES at the same time. For single-channel second-order attack, we first calculate the absolute difference between Sbox position $i$ and $j$ in power/EM traces ($P_i$,$P_j$,$EM_i$,$EM_j$) to get power absolute difference $AP_{ij}$ and EM absolute difference $AEM_{ij}$ in Eqns.~\eqref{formula:AP} and~\ref{formula:AEM}. The power and EM absolute value is adopted to describe $||HW(sbox'(state[i])|-|HW(sbox'(state[j]))||$ for single-channel power/EM attacks.
\begin{equation}
AP_{ij}= ||P_i|-|P_j||
\label{formula:AP}
\end{equation}
\begin{equation}
AEM_{ij}= ||P_i|-|P_j||
\label{formula:AEM}
\end{equation}

For the dual-channel (power \textit{and} EM) second-order attack, we adopt combination coefficient $\alpha$ (range in [0,1]) to combine absolute power value($AP_{ij}$) and absolute EM value($AEM_{ij}$) to get absolute combined value $AZ_{ij}$\cite{bai2023dual}.
\begin{equation}
AZ_{ij}= \alpha*AP_{ij}+(1-\alpha)*AEM_{ij}
\label{formula:combine-2nd1}
\end{equation}
Then, we calculate the index of combination coefficient $\alpha$ using the method proposed in our prior work\cite{bai2023dual} that makes the highest correlation $I$ between absolute combined value $AZ_{ij}$ and two Sbox output hypothesis $HW(sbox'(state[i]) \oplus sbox'(state[j]))$.

\subsection{SPERO Dataset}
In this section, we briefly describe our SPERO dataset, including the unmasked/masked AES code, our experimental setup, and the structure of our dataset.

\subsubsection{AES code}
There are two AES codes considered in this paper, and a brief explanation of each is given below. \textbf{Unmasked AES} is typical AES-128 (in ECB mode) software implementation and it could be found on website\cite{AEScode}. \textbf{Masked AES} is from Github\cite{maskedAES}, which is used by many peers in this area. The masked AES code generates the mask at the beginning of the round and removes them after the \textit{ShiftRows} operation. Thus, the selected masked AES code does not leak during the Sbox computation.

\subsubsection{Experiment Setup}
In our experiments, we gather EM and power traces at the same time when the attack target is encrypting data. The target board is Arduino UNO and its core frequency is 16MHz. The sampling speed of our oscilloscope MDO3102 is set to 100MS/s in the unmasked AES experiment and 500MS/s in the masked AES experiment. The sampling speed of RASCv3 is 160MS/s in both masked/unmasked experiment. 

\subsubsection{Dataset Structure}

\begin{figure}
  \centering    \includegraphics[width=0.29\textwidth]{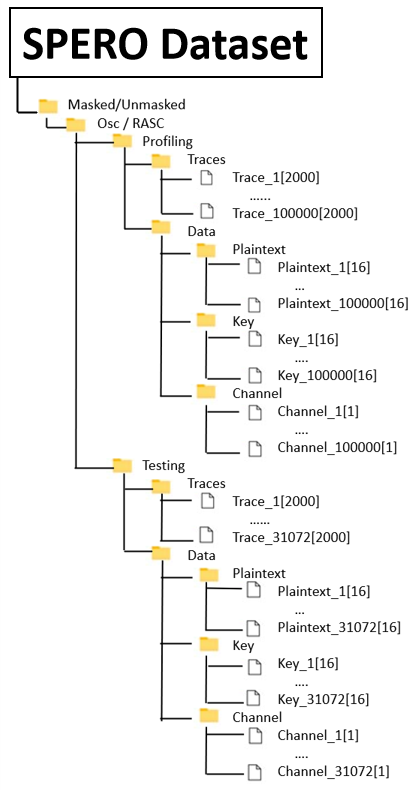}
  
     \caption {Unmasked/masked AES dataset organization. }
     \label{dataset}
\end{figure}

Our dataset\cite{SPERO} follows the format and structure of the popular side-channel attack dataset called ASCAD\cite{rioja2021auto, ANSSI, AESPT, benadjila2020deep}. As shown in Figure~\ref{dataset}, we collected 2000-feature power/EM traces for the third subkey of the first round of unmasked and masked AES encryptions. Since each subkey of unmasked/masked AES consists of 8 bits, there are a total of 256 possible subkey values. For each subkey value, we have gathered 256 traces, where the plaintext at position 3 varies from 0x00 to 0xFF. Consequently, there are a total of 131072 (256$\times$256$\times$2) power/EM traces for both unmasked and masked AES encryption. In these 131072 traces, the first 100000 of EM/power traces have been organized into the profiling folder, while the remaining 31072 traces have been placed in the testing folder. Furthermore, each power/EM trace is labeled sequentially and accompanied by essential information, including the label's sequence number, channel, and plaintext value. Lastly, we have included a tutorial text to guide users in opening the files using the provided Python script. Upon publication, the unmasked/masked AES ASCAD file will be made available for download on GitHub.

\subsection{Single and Dual-channel Attack Results on Unmasked AES}

In this subsection, we examine the measurements-to-disclosure (MTD) for key extraction against unmasked AES in both offline and real-time modes. The offline mode result is presented in Table~\ref{table:inPCunmasked} and it could be found in our earlier work\cite{bai2023dual}. For offline mode, Table~\ref{table:inPCunmasked} shows the MTD for 100\% success rate with the differential attack for all 16 AES-128 subkeys on two Arduino UNO boards. We profile on Board 1 and test on Boards 1 and 2 to examine generalizability. To be more specific, power/EM channel refers to only using power/EM features and the ``Combined'' results combine EM and power channel using the dual channel methodology presented in\cite{bai2023dual}. In the offline experiments, we arrange the commercial EM probe closely near the signal source to collect EM traces. For collecting EM waves, the selected position is closer to the signal source and the interference signals/white noise on board attenuate through the air. Thus, as presented in a table, the power channel has a lower SNR ratio than the EM channel and needs more traces to achieve a full extraction rate than the EM channel. Besides, the experimental result in the table shows the combined feature is more efficient in extracting the secret key. That is, the proposed approach only needs 267 traces on average for extracting subkeys from board 1 and 272 traces for board 2.

\begin{table}[t]
 \scriptsize
 \centering
 \caption{MTD for unmasked AES-128 encryption module in offline mode using differential attack.}
\begin{tabular}{|c|c|c|c|c|c|c|}
\hline
\multirow{2}{*}{Subkey} & 
\multicolumn{2}{c|}{Power}  &
\multicolumn{2}{c|}{EM}  &
\multicolumn{2}{c|}{Combined}  
\\ \cline{2-7} & Board 1 & Board 2  &  Board 1 & Board 2 &  Board 1 & Board 2 

\\ \hline
1 & 150 & 150  & 100 & 100 & 70 & 85
\\ \hline
2 & 400 & 350 & 250 & 250 & 160 & 200
\\ \hline
3 & 950 & 900 & 750 & 600 & 420 & 400
\\ \hline
4 & 800 & 800 & 400 & 400 & 350 & 350
\\ \hline
5 & 800 & 800  & 380 & 400 & 320 &  300
\\ \hline
6 & 200 & 170  & 80 & 100 & 40 & 50
\\ \hline
7 & 1200 & 1150 & 1000 & 950 & 450 & 450
\\ \hline
8 & 1000 & 1000 & 700 & 750 & 400 & 420
\\ \hline
9 & 800 & 800  & 400 & 380 & 300 & 300
\\ \hline
10 & 200 & 250 & 150 & 150 & 60 & 50
\\ \hline
11 & 300 & 320 & 250 & 280 & 150 & 180
\\ \hline
12 & 1000 & 900 & 650 & 600 & 450 & 430
\\ \hline
13 & 1000 & 950 & 550 & 550 & 400 & 400
\\ \hline
14 & 300 & 320 & 200 & 200 & 140 & 150
\\ \hline
15 & 700 & 650  & 400 & 450 & 350 & 350
\\ \hline
16 & 400 & 400 & 250 & 250 & 200 & 250
\\ \hline \hline
Avg. &637  &619  &394 &400 &267 & 272
\\ \hline
\end{tabular}
 \label{table:inPCunmasked}
\end{table}

In the real-time mode, we implement DPA algorithms onto RASC and let it extract the subkey bits internally. Before that, we adopt feature key extraction methods such as minimum redundancy and max relevance (mRMR~\cite{peng2005feature}) to determine the useful features. This is significantly important for better combining the power/EM features and saving the FPGA's memory. The details of how we implement the proposed methodology in real-time were presented in\cite{bai2023dual}. We perform real-time attacks using RASCv2 and RASCv3 and report the results in Tables~\ref{table:inRASCunmaskedv2}and~\ref{table:inRASCunmaskedv3}, respectively. Compared with the offline mode results in Table~\ref{table:inPCunmasked}, the real-time mode results need more traces to achieve a 100\% success rate. The average MTD for power/EM/combined channel of board 1 using RASCv2 is 3171, 5637, and 2450. Using RASCv3, they are 2571, 4040, and 2134, demonstrating that it needs fewer traces due to its higher sensitivity (12-bit vs. 8-bit). Besides, the LT2242\cite{LT2242} ADC on RASCv3 has advantages in lower SNR(65.4dB) compared with ADC08200 ADC(43.4dB) on RASCv2. Furthermore, the detecting range of the ADC08200 ADC on the RASCv2 board is set to the range of 0V to 3V\cite{bai2022rascv2} and this causes some issues when detecting minus voltages. Though the external voltage level shifter circuit could raise trace voltage to 0V--3V for RASCv2, it still affects lowering the SNR ratio of the whole EM traces, and this is part of the reason that RASCv3 uses fewer MTD EM traces than RASCv2 in real-time mode. Another reason contributes to the usage of various antennae for RASCv3. In the real-time mode of attacking unmasked AES inside RASCv2, the internal EM antenna inside RASCv2 is distributed in four inner layers of RASC. The bad interconnection between different antenna layers undermines the SNR of collected EM traces from RASCv2. For RASCv3, we adopt a printable antenna to receive EM waves. As shown in Figure~\ref{magnitude_comparison}, the printable antenna for RASCv3 has a higher magnitude than the internal antenna inside RASCv2 and a better structure in design. This leads to a higher SNR ratio of the EM waves collected by RASCv3, and less MTD to achieve 100\% subkey extracting rate in the EM channel. 

Compared with the offline mode results, RASCv3 still needs more traces to achieve a 100\% success rate, either in single Powe/EM or dual-channel. Part of the reason is we bypass division and decimal calculation inside FPGA, and this reduces the attack accuracy. Other potential reasons are due to the BNC connector port and better impedance matching in the oscilloscope. This could always let oscilloscopes achieve better sampling in collecting EM or power traces. Nevertheless, the real-time experiments still demonstrate the efficiency of the proposed dual-channel methodology in paper\cite{bai2022real}. That is, it achieves a higher success rate compared to a single EM/power channel with the same amount of traces and has lower MTD.

 \begin{table}[t]
 \scriptsize
 \centering
 \caption{MTD for unmasked AES-128 encryption module in real-time using differential attack inside RASCv2\cite{bai2023dual}. }
\begin{tabular}{|c|c|c|c|c|c|c|}
\hline
\multirow{2}{*}{Subkey} & 
\multicolumn{2}{c|}{Power}  &
\multicolumn{2}{c|}{EM}  &
\multicolumn{2}{c|}{Combined}  
\\ \cline{2-7} & Board 1 & Board 2  &  Board 1 & Board 2 &  Board 1 & Board 2 

\\ \hline
1 & 350  & 500  &1200  & 1200 &300  & 400
\\ \hline
2 & 2500 & 3000 & 6000  & 6500 &2000  & 2200
\\ \hline
3 & 5000 & 5500 &7000  & 7500 &4000  & 4500
\\ \hline
4 & 4000 & 5000 &7500  & 7800 & 3500 & 4000
\\ \hline
5 & 3000 & 5000 &5000  & 6000 & 2500 & 3500
\\ \hline
6 &150  & 200  &500  & 650 &100  & 150
\\ \hline
7 & 5500 & 6000 & 10000 & 10000 &4000  & 4500
\\ \hline
8 &2500  & 3500 & 5500 & 6500 & 2000 & 3000
\\ \hline
 9 &3000  & 3500 & 8000 & 8000 &2000  & 3000
\\ \hline
10 & 550 & 900  &3000  & 4000 & 500 & 700
\\ \hline
 11 & 3000 & 4000  &7500  & 8000 & 2500 & 3000 
\\ \hline
12 & 6000 & 6500 & 9000 & 9000 & 4000 & 4200 
\\ \hline
13 & 6000 & 6500  &9500  & 9500 &4500  & 4700
\\ \hline
14 & 3000 & 3000 & 7000 & 7000 & 2500 & 2500
\\ \hline
 15 & 5000 & 5000 &1000  & 10000 & 4000 & 4000
\\ \hline
16 & 1200 & 1500 &2500  & 2500 &800  & 900
\\ \hline \hline
Avg. &3171 &3725 &5637 & 6509 &2450 &2828
\\ \hline

\end{tabular}
 \label{table:inRASCunmaskedv2}
\end{table}

 \begin{table}[t]
 \scriptsize
 \centering
 \caption{MTD for unmasked AES-128 encryption module in real-time using differential attack inside RASCv3. } 
\begin{tabular}{|c|c|c|c|c|c|c|}
\hline
\multirow{2}{*}{Subkey} & 
\multicolumn{2}{c|}{Power}  &
\multicolumn{2}{c|}{EM}  &
\multicolumn{2}{c|}{Combined}  
\\ \cline{2-7} & Board 1 & Board 2  &  Board 1 & Board 2 &  Board 1 & Board 2 

\\ \hline
1 &300  &350   &650  &800  & 200&250 
\\ \hline
2 &2200  &2300  &4000 &5000 & 1800 &1900 
\\ \hline
3 &4500  &4600  &6000 & 7000 & 3600 & 3800
\\ \hline
4 &3500  &3800  & 5000 &6000& 3000 & 3100
\\ \hline
5 & 2500 &2700  &4000 & 4600 & 2000 & 2300
\\ \hline
6 & 150 &200 &500 &600  &100&150 
\\ \hline
7 &5000  &5300  & 7000&8000  & 3500 & 3900
\\ \hline
8 &2000  &2200  &3500 &4500  & 1700 & 1900
\\ \hline
 9 &2000  &2400  &5000 &6000  & 1500 &1800 
\\ \hline
10 &500  & 800  &2000 &2500  & 450  & 550
\\ \hline
 11 &2500  & 2800  & 3500&4000  & 2300 & 2800 
\\ \hline
12 &4500  & 4800  & 5500 & 6500 & 4000 & 4300 
\\ \hline
13 &4500  & 5000  & 6000 & 7200 & 4000 & 4200 
\\ \hline
14 &2500  & 2700 & 4000 & 5300 & 2200 & 2300
\\ \hline
 15 &3500  & 4000 & 6000 & 7000 & 3000 & 3200
\\ \hline
16 &1000  & 1300 & 2000 & 2500 & 800 & 900
\\ \hline \hline
Avg. &2571 &2828 &4040 & 4843 &2134 & 2333
\\ \hline

\end{tabular}
 \label{table:inRASCunmaskedv3}
\end{table}

\subsection{Single and Dual-channel Attack Results on Masked AES}

\begin{figure}
  \centering
    \includegraphics[width=0.4\textwidth]{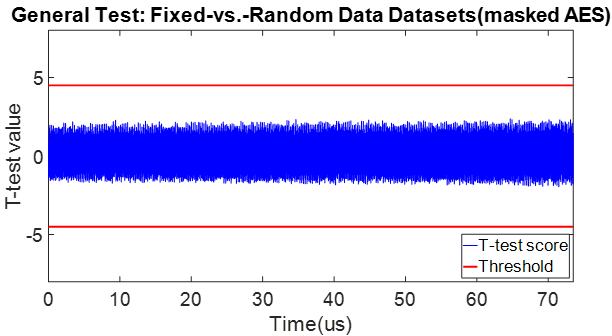}
     \caption {T-test result of masked AES using 10M traces.}
     \label{T_Test}
\end{figure}

The masked AES generates random masks during each encryption, making a first-order SCA theoretically impossible. Here, we collect 10 million traces during a fixed vs random data T-test~\cite{TVLA} to demonstrate the robustness of the masked implementation. For the random dataset, the first plaintext is 0xAAAAAAAAAAAAAAAA. Then, the ciphertext is used to be the next plaintext. For the fixed dataset, the first plaintext is fixed to 0xAAAAAAAAAAAAAAAA. After gathering all the traces, we use the Python code from~\cite{ttestcode} to calculate the $t$-test result. Even after collecting 10 million traces, the t-test value remains under the threshold ($\pm4.5\sigma$) as shown in Figure~\ref{T_Test} and is thus considered secure against first-order attacks.

For extracting subkeys from masked AES, we use the second-order attack described in Section~\ref{sec:2orderattack} to extract two subkeys at the same time, e.g., subkeys 1 and 2, subkeys 3 and 4, $\hdots$, subkeys 15 and 16. We list the second-order differential attack MTD against masked AES offline mode using traces collected by an oscilloscope in Table~\ref{table:offlinemaskedosc} and RASCv3 in Table~\ref{table:offlinemaskedRASC}. For this comparison, we arrange the printable antenna near the signal source to collect EM traces in both oscilloscope and RASC experiments. In this section, RASCv3's FPGA cannot satisfy the memory requirements of the second-order attack, and thus we cannot perform its attack in real-time mode. In next version of RASC, a larger memory may be included to address this.

Using an oscilloscope with RASC's printable antenna, the EM channel in Table~\ref{table:offlinemaskedRASC} needs 2.5$\times$ more traces than the power channel. However, in the case of unmasked AES in Table~\ref{table:inRASCunmaskedv3}, the EM traces are also collected with a printable antenna but only need 70\% traces more than the power channel. Unlike the first-order attack, the second-order attack utilizes the differences of trace sections which execute the same instructions but from different Sbox operations. For getting the correct trace difference in the second order attack, the trace segment of the collected trace should not only consider the accuracy among other collected traces but also need to consider the accuracy versus other segments in the same traces. To be more specific, if we stack all traces to a matrix, the first-order attack only needs to consider the detection accuracy in vertical direction but the second-order attack needs to consider the accuracy in both vertical and horizontal directions. The higher amplitude of traces makes it easier to get differences in both vertical and horizontal directions and could be beneficial to the subkey extraction efficiency. 

When RASC is used instead on an oscilloscope, we need 128500 power traces, 307375 EM traces, and 111875 dual-channel traces on average to extract subkeys. 
The main reason that RASC needs more EM/power traces is its lower sampling speed, lower sensitivity, and reduced tolerance to clock jitter. 
The former two have already been discussed in Section~\ref{sec:RASCvScope}.
In the case of the latter, 
clock jitter is also important to the accuracy of collecting traces. The oscilloscope's jitter ($<$ 0.01\%) is two orders of magnitude less than RASC ($\approx$1\%). A high clock jitter could cause a mismatch when we subtract two Sbox operations in the traces during the second order attack. Nevertheless, the combined channel still uses 13\% less traces to achieve a 100\% rate versus the power channel and 63\% less than the EM channel.

As for the comparison of MTDs in unmasked and masked AES, the masked AES needs many more traces to achieve a 100\% cracking rate than the unmasked case. Besides the challenges in trace amplitude, the second-order attack also needs to consider DC shift in both vertical and horizontal directions. On the other hand, the first-order attack only needs to consider the DC shift in the vertical direction. Besides, the second-order attack requires more hypotheses (65536) than the first-order (256), and this increases the possibility of failure in the subkey extraction. The combined channel still uses 10\% less traces to achieve a 100\% rate versus power channel and 75\% less than EM channel. This demonstrates the benefits of the proposed combination methodology not only in unmasked AES but also against masked AES.

 \begin{table}[t]
 \scriptsize
 \centering
 \caption{MTD for masked AES-128 encryption module in offline mode oscilloscope using 2nd-order differential attack. } 
\begin{tabular}{|c|c|c|c|c|c|c|}
\hline
\multirow{2}{*}{Subkey} & 
\multicolumn{2}{c|}{Power}  &
\multicolumn{2}{c|}{EM}  &
\multicolumn{2}{c|}{Combined}  
\\ \cline{2-7} & Board 1 & Board 2  &  Board 1 & Board 2 &  Board 1 & Board 2 
\\ \hline
1 & 21000 & 22000  & 65000 & 70000 &19500 & 20000
\\ \hline
2 & 21000  & 22000 &65000 & 70000 & 19500 & 20000
\\ \hline
3 &  20000 & 23000  & 67000 & 69000 &18000  & 19000
\\ \hline
4 & 20000  & 23000 & 67000& 69000  &18000  & 19000
\\ \hline
5 & 18000  & 20000 & 63000& 66000 &16500  & 17500
\\ \hline
6 & 18000 & 20000 & 63000& 66000 &16500  & 17500
\\ \hline
7 & 23000 & 23000  &65000 & 66000 &20000  & 20000
\\ \hline
8 & 23000 & 23000 &65000 & 66000 &20000  & 20000
\\ \hline
9 & 17500 & 19000 &70000 & 70000 &15500  & 16000
\\ \hline
10 & 17500 & 19000 &70000 & 70000 &15500  & 16000
\\ \hline
11 & 20000 & 20000 &68000 & 70000 &18000  & 18000
\\ \hline
12 & 20000 & 20000 &68000 & 70000 &18000  & 18000
\\ \hline
13 & 18500 & 19000 & 72000& 75000 &16000  & 17000
\\ \hline
14 & 18500 & 19000 & 72000& 75000 &16000  & 17000
\\ \hline
15 & 19000 & 21000 &73000 & 75000 &17000  & 18500
\\ \hline
16 & 19000 & 21000 &73000 & 75000 &17000  & 18500
\\ \hline \hline
Avg. &19625 & 20875& 67875&70125  &17563 & 18250
\\ \hline
\end{tabular}
 \label{table:offlinemaskedosc}
\end{table}

 \begin{table}[t]
 \scriptsize
 \centering
 \caption{MTD for masked AES-128 encryption module in offline mode RASC using 2nd-order differential attack. } 
\begin{tabular}{|c|c|c|c|c|c|c|}
\hline
\multirow{2}{*}{Subkey} & 
\multicolumn{2}{c|}{Power}  &
\multicolumn{2}{c|}{EM}  &
\multicolumn{2}{c|}{Combined}  
\\ \cline{2-7} & Board 1 & Board 2  &  Board 1 & Board 2 &  Board 1 & Board 2 
\\ \hline
1 &120000  &140000   &300000 &320000  &100000 &110000 
\\ \hline
2 &120000  &140000  &300000 &320000 &100000  &110000 
\\ \hline
3 &130000  &140000  &320000 &340000  &110000  &120000 
\\ \hline
4 &130000  &140000  &320000 & 340000 &110000  &120000
\\ \hline
5 &118000  &130000  &290000 & 300000 &100000  & 110000
\\ \hline
6 &118000  &130000  &290000 &300000  &100000  & 110000
\\ \hline
7 &125000  &140000  &310000 &330000  &110000  & 115000
\\ \hline
8 &125000  &140000  &310000 & 330000 &110000 & 115000 
\\ \hline
9 &140000  &140000  &330000 & 370000 &130000  & 135000
\\ \hline
10 &140000  &140000  &330000 &370000  &130000  & 135000 
\\ \hline
11 &138000  &145000  &319000 &350000  &120000  & 130000 
\\ \hline
12 &138000  &145000  &319000 &350000  &120000  & 130000
\\ \hline
13 &135000  &150000  &290000 &310000  &120000  & 130000
\\ \hline
14 &135000  &150000  &290000 & 310000 &120000  & 130000
\\ \hline
15 &122000  & 135000 &300000 & 310000 &105000  & 115000
\\ \hline
16 &122000  &135000  &300000 & 310000 &105000  & 115000
\\ \hline \hline
Avg. &128500 &140000 &307375 &328750  &111875 &120568
\\ \hline
\end{tabular}
 \label{table:offlinemaskedRASC}
\end{table}

\section{Conclusion and future work}\label{sec:conclusion}

In this paper, we successfully demonstrate the attack capability of RASCv3 in both masked and unmasked AES subkey extraction experiments. Compared with RASCv2, RASCv3 upgrades its functionality in sampling speed, sensitivity, and EM wave detection. In the unmasked AES subkey extraction experiment, RASCv3 utilizes fewer traces to achieve 100\% subkey extraction rate than RASCv2. Compared with other side-channel instruments, RASCv3 has advantages in its low price, small size, and portable features, such as remote communication. In the masked AES subkey extraction experiment, we successfully combine dual-channel traces to extract the subkey from masked AES and the experiment result shows 10\% and 75\% fewer traces are needed in dual channel than EM/power channel with oscilloscope and 13\% and 63\% fewer traces are needed in the dual channel than EM/power channel with RASCv3. Besides, we generate the SPERO dataset by gathering EM/Power channel traces of unmasked/masked AES at the same time during encryption and make it available to the community.

In future work, we will upgrade RASCv3 to a new version that consists of a larger memory FPGA, and ADCs with higher sampling rates. With the improved version of RASCv3, we can sample more accurately and increase the accuracy of collecting traces in masked AES subkey extraction. Besides, larger memory could contain more hypotheses and implement the second-order attack algorithm in real-time.

\balance

\bibliographystyle{IEEEtran}
\bibliography{reference/references}

\vspace{12pt}
\color{red}

\end{document}